\documentclass[conference]{IEEEtran}
\IEEEoverridecommandlockouts
\usepackage{amsmath,amssymb,amsfonts}
\usepackage{algorithmic}
\usepackage{graphicx}
\usepackage{textcomp}
\usepackage{xcolor}
\usepackage{bbm}
\usepackage{algorithm}
\usepackage{flushend}
\usepackage[numbers]{natbib}

\usepackage{subfig}
\usepackage{svg}

\def\BibTeX{{\rm B\kern-.05em{\sc i\kern-.025em b}\kern-.08em
    T\kern-.1667em\lower.7ex\hbox{E}\kern-.125emX}}
\begin{document}

\title{GraphCue for SDN Configuration Code Synthesis\\
}

\author{
    \IEEEauthorblockN{Haomin Qi$^{1}$, Fengfei Yu$^{1}$, Chengbo Huang$^{2}$}
    \IEEEauthorblockA{$^1$University of California San Diego, La Jolla, CA, USA \\ $^2$Columbia University, New York City, NY, USA \\ Email: \{h5qi, fey006\}@ucsd.edu, ch4019@columbia.edu}
    \vspace{-2.5em}
}

\maketitle

\begin{abstract}
We present \textit{GraphCue}, a topology-grounded retrieval and agent-in-the-loop framework for automated SDN configuration. Each case is abstracted to a JSON graph and embedded by a lightweight three-layer GCN trained with contrastive learning. The nearest validated reference is injected into a structured prompt that constrains code generation, and a verifier closes the loop by executing the candidate and feeding failures back. On 628 validation cases, GraphCue reaches 88.2\% pass within 20 iterations and completes 95\% of verification loops within 9\,s. Ablations without retrieval or prompt structure are substantially weaker, indicating that topology-aware retrieval and constraint conditioning are key drivers of performance.
\end{abstract}


\section{Introduction}

Production SDN deployments span hundreds of devices and heterogeneous protocol stacks, and manual scripting becomes brittle at this scale. Pure text generation ignores topology, while static checks miss run-time semantics \cite{falkner2022ibn}. Operators need a method that aligns configuration synthesis with network structure and that verifies behavior as part of generation.

\textbf{GraphCue} addresses this need with three components. First, source artifacts are mapped to JSON graphs with node features and preserved link multiplicity. A compact GCN encoder produces graph-level embeddings and enables nearest-neighbor retrieval among validated cases. Second, a retrieval-conditioned prompt composes the target graph, selected reference snippets, concise schema knowledge, and checkable constraints to ground code synthesis. Third, an LLM agent interacts with a containerized verifier that executes candidates and returns machine-readable reports; the agent refines prompts using these reports and iterates until pass or budget. Our method, GraphCue, achieves an 88.2\% pass rate within 20 iterations and short per-loop latencies.


\section{Methdology}

\subsection{Data and JSON Graph Abstraction}

We use the open FRRouting corpus as ground truth. After de-duplication, there are 4{,}128 cases with a topology description, a Python driver, and, when applicable, a P4 program. The split is 3{,}500 training and 628 validation cases, disjoint at case and file level.

\begin{figure}[t]
  \centering
  \includegraphics[width=\linewidth]{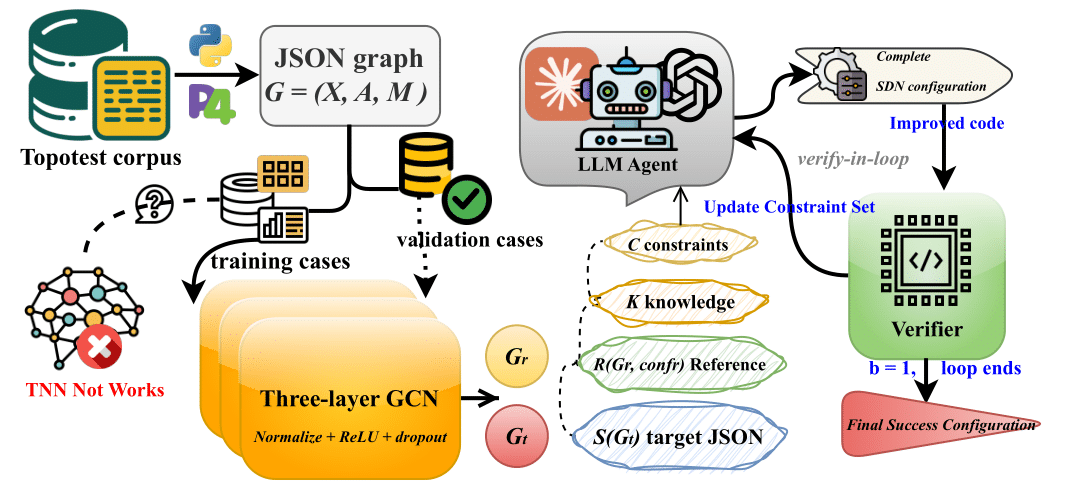}
  \caption{GraphCue pipeline overview}
  \label{fig:vali}
  \vspace{-1.5em}
\end{figure}

A parser \(\Phi\) extracts devices \(V\), interfaces \(I\), and a link multiset \(E\). Each case maps to a JSON graph \(G=(X,A,M)\) where \(X\in\mathbb{R}^{N\times F}\) holds node features, \(A\in\{0,1\}^{N\times N}\) is an undirected adjacency with self-loops, and \(M:V\times V\rightarrow\mathbb{N}\) preserves parallel links. Features cover device type, degree and light structure, protocol and intent counters, and hashed tokens for interface names and addressing. Connectivity statements induce edges in \(E\). Because \(M\) is explicit, the mapping is lossless for adjacency, endpoints, and multiplicity while discarding control flow and retaining only relations that affect connectivity. The full flow appears in Fig.~\ref{fig:vali}.

\subsection{Network Training and GraphCue}

We first evaluate topological neural networks (TNNs) on cell and simplex representations. We use the convention that $B_0\in\mathbb{R}^{|V|\times|E|}$ is the node–edge incidence and $B_1\in\mathbb{R}^{|E|\times|F_2|}$ is the edge–face incidence, with Hodge Laplacians $L_k=B_k^{\top}B_k+B_{k+1}B_{k+1}^{\top}$. In our corpus, valid two-cells are rare, so $B_1\approx 0$ and the edge-level Laplacian reduces to $L_1\approx B_0^{\top}B_0$, which removes curvature cues and leads to collapsed or unstable training \cite{bodnar2021cell}. A transformer with Laplacian positional encodings also underperforms due to small graph diameters and attention concentrated on hubs.

We adopt a three-layer GCN with global pooling to obtain 32-dimensional graph embeddings from $(X,A)$ \cite{kipf2017gcn}. Let $f_{\theta}$ produce node embeddings $H=f_{\theta}(X,A)\in\mathbb{R}^{N\times d}$ and let $p(\cdot)$ be mean pooling. The graph vector is $\tilde z=p(H)\in\mathbb{R}^{d}$, and the embedding used for retrieval is the $\ell_2$-normalized $z=\tilde z/\lVert \tilde z\rVert_2\in\mathbb{R}^{32}$. Training is contrastive \cite{you2020graphcl}: two independent augmentations $\mathcal{T}_1,\mathcal{T}_2$ drop edges with probability $0.2$ and nodes with probability $0.1$. For sample $i$ we set $z_i^{(1)}=z(\mathcal{T}_1(G_i))$ and $z_i^{(2)}=z(\mathcal{T}_2(G_i))$ and minimize the InfoNCE loss with temperature $\tau=0.2$ and batch size $B$,
\[
\mathcal{L}=-\frac{1}{B}\sum_{i=1}^{B}\log\frac{\exp\!\big(\langle z_i^{(1)},z_i^{(2)}\rangle/\tau\big)}{\sum_{j\neq i}\exp\!\big(\langle z_i^{(1)},z_j^{(2)}\rangle/\tau\big)},
\]
where $\langle\cdot,\cdot\rangle$ is the inner product, equal to cosine similarity under normalization. We train for 80 epochs with Adam at learning rate $10^{-4}$. A single 16\,GB GPU suffices, and the code also runs on a CPU host with a longer runtime. After convergence the encoder is frozen. Cosine retrieval over the reference set selects one neighbor per query. Similarity heatmaps are block-diagonal, and nearest pairs above 0.90 match manual checks of device counts, degree profiles, and interface attachments.

GraphCue prompting explicitly composes $\mathcal{P}=[\mathcal{S}(G_t);\mathcal{R}(G_r,\mathrm{conf}_r);\mathcal{K};\mathcal{C}]$, where $\mathcal{S}$ serializes the target JSON, $\mathcal{R}$ adds the retrieved JSON and selected configuration snippets, $\mathcal{K}$ supplies compact background knowledge, and $\mathcal{C}$ states checkable constraints on interface attachment, addressing, naming, and daemon activation. The model returns a main driver and per-device configuration files conditioned on $\mathcal{P}$.

\subsection{LLM Agent in the Verification Loop}

An LLM agent coordinates end-to-end synthesis and interacts with a containerized verifier through a narrow API. The verifier provisions the target topology from the JSON graph, applies a candidate configuration, executes checks, and returns a machine-readable report.

Given a target graph \(G\) and a candidate \(C\), the verifier returns a pass flag \(b\in\{0,1\}\) and a report \(R\) that includes interface state, adjacency formation per protocol, reachability between selected endpoints, addressing and naming invariants, and trimmed logs. Checks obey timeouts and resource limits. Provisioning is deterministic under fixed seeds. Execution is idempotent; a failed attempt triggers cleanup and reprovisioning.

The agent runs a verify-in-loop process. At iteration \(t\) it forms a prompt from the target serialization, the retrieved materials, a compact background block, and a constraint set updated from the previous report. Missing interfaces become explicit attachment directives, adjacency mismatches become neighbor definitions with required parameters, and reachability failures become addressing fixes or redistribution rules tied to the affected subgraph. The agent queries the model, submits the new candidate to the verifier, reads the flag and the report, and refines the next prompt. The loop stops with success when any iteration returns \(b=1\) and outputs the accepted configuration. It stops with failure when a fixed iteration budget \(T\) or a wall-clock limit is reached; the system then emits the best attempt, the final report, and all archived artifacts.

\vspace{-1em}
\section{Evaluation}

\begin{figure}[t]
  \centering
  \includegraphics[width=\linewidth]{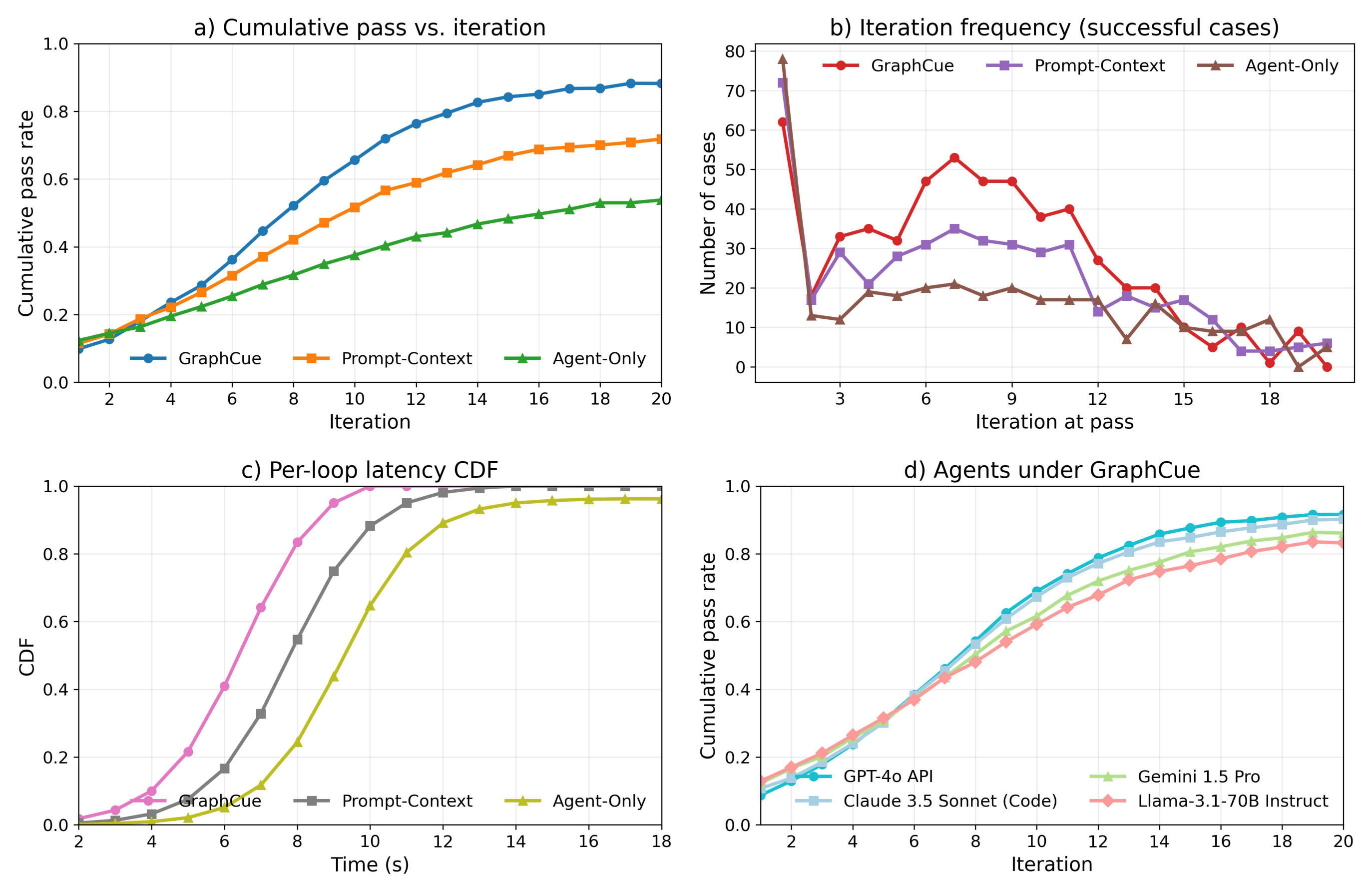}
  \caption{Experimental results in four aspects.}
  \label{fig:exp}
  \vspace{-1.5em}
\end{figure}

\textbf{Setup and views.} We evaluate three groups on the 628-case validation split with a verify-in-loop budget of 20 iterations and measure per-loop wall-clock latency that includes one agent invocation and one full verification cycle. The three groups are (1) GraphCue, (2) Prompt-Context without retrieval, and (3) Agent-Only without prompt scaffolding. Panels (a)–(c) in Fig.~\ref{fig:exp} use Claude 3.5 Sonnet (Code) as the LLM agent for all three groups. Panel (d) fixes GraphCue and compares four agents: GPT-4o API, Claude 3.5 Sonnet (Code), Gemini 1.5 Pro, and Llama-3.1-70B Instruct. The four reported views correspond to panels (a)–(d), respectively: cumulative pass rate by iteration, distribution of iterations required to pass, empirical cumulative distribution function of per-loop latency, and cumulative pass rate by iteration across the four agents. Here the CDF reports the fraction of loops that complete within a given time \(t\).

\textbf{Results.} In Fig.~\ref{fig:exp}(a) cumulative pass rates rise with iteration: GraphCue reaches 88.2\% by iteration 20, Prompt-Context 71.8\%, and Agent-Only 53.8\%. By iteration 5 the rates are 54.0\%, 38.0\%, and 25.2\%. Fig.~\ref{fig:exp}(b) shows that among passing cases, 36\% of GraphCue successes finish by iteration 5 and 72\% by iteration 10, compared with 20\% and 51\% for Prompt-Context and 12\% and 34\% for Agent-Only, for totals of 554, 451, and 338 passes. In Fig.~\ref{fig:exp}(c) the latency CDFs indicate that 95\% of loops finish within 9, 11, and 14 seconds for the three groups, with medians of 6.5, 7.8, and 9.2 seconds. Fig.~\ref{fig:exp}(d) shows agent sensitivity under GraphCue: GPT-4o API reaches 91.6\% by iteration 20, Claude 90.2\%, Gemini 86.1\%, and Llama 83.2\%; early gains at iteration 5 are 58.6\% and 56.9\% for GPT-4o API and Claude.

\vspace{-0.5em}
\section{Conclusion}
We introduce \textbf{GraphCue}, a topology-aware retrieval-conditioned prompting and agent-in-the-loop verification framework for SDN configuration synthesis. On 628 validation cases, GraphCue attains an 88.2\% pass rate within 20 iterations. Future work includes safety-oriented distributed and edge deployment, pre-training and domain adaptation of the encoder and the language model \cite{qi2025aba}, and richer verification with invariants and fault injection.



\end{document}